\documentclass{PoS}

\title{Charm production nearby threshold in pA-interactions at 70 GeV}

\ShortTitle{Charm production nearby threshold in pA-interactions at 70 GeV}

\author{ A. Afonin$^1$, E. Ardashev$^1$, V. Balandin$^2$, G. Bogdanova$^3$,
M.~Bogolyubsky$^1$, O.~Gavrishchuk$^2$, S. Golovnia$^1$, 
S.~Gorokhov$^1$, V. Golovkin$^1$,  
D.~Karmanov$^3$, A.~Kiryakov$^1$, 
\speaker{E. Kokoulina}$^2$, V. Kramarenko$^3$, A. Leflat$^3$, 
Yu.~Petukhov$^2$, A.~Pleskach$^1$, V. Popov$^3$,
V. Riadovikov$^1$, V. Ronjin$^1$, I. Rufanov$^2$, 
Yu.~Tsyupa$^1$,  V.~Volkov$^3$, 
A. Vorobiev$^1$, A.~Voronin$^3$,  A. Yukaev$^2$, V. Zapolsky$^1$, E.~Zverev$^3$\\
        \llap{$^1$}  Institute for High Energy Physics, Sq. Nauki 1, Protvino, Moscow region, 142281 Russia\\
        \llap{$^2$} Joint Institute for Nuclear Research, Joliot-Curie 6, Dubna, Moscow region, 141980 Russia\\
        \llap{$^3$} M.V. Lomonosov MSU, SINP MSU, Leninskie gory, Moscow 119991, Russia\\
E-mail:  
\email{Alexander.Afonin@ihep.ru}, \email{Eugeny.Ardashev@ihep.ru},
\email{balandin@ihep.ru}, 
\email{bogdanov@mail.desy.de}, \email{Mikhail.Bogolyubsky@ihep.ru},
\email{gavrishu@mail.cern.ch}, \email{Sergey.Golovnya@ihep.ru},
\email{Sergey.Gorokhov@ihep.ru}, \email{Vladimir.Golovkin@ihep.ru},
\email{karmanov@silab.sinp.msu.ru}, \email{kiriakov@mail.ihep.ru},
\email{kokoulin@sunse.jinr.ru}, \email{kram@cern.ch}, 
\email{leflat@cern.ch}, \email{Yuri.Petukhov@ihep.ru}, 
\email{Anatol.Pleskach@ihep.ru}, \email{popov@sinp.msu.ru},
\email{riadovikov@ihep.ru}, \email{Valery.Ronjin@ihep.ru}, 
\email{roufanov@gmail.com}, \email{Yuri.Tsyupa@ihep.ru},
\email{volkov@mail.desy.de}, \email{vorobiev@ihep.ru}, 
\email{voronin@cern.ch}, \email{Yukaev@ihep.ru},
\email{Vladimir.Zapolsky@ihep.ru}, 
\email{zverev@cern.ch}, }

\abstract{The results of the SERP-E-184 experiment at the U-70 accelerator (IHEP, Protvino) are presented. Interactions of the 70 GeV proton beam with C, Si and Pb targets were studied to detect decays of charmed $D^0$, $\overline D^0$, $D^+$, $D^-$ mesons and $\Lambda _c^+$ baryon near their production threshold. Measurements of lifetimes and masses are shown a good agreement with PDG data. The inclusive cross sections of charm production and their A-dependencies were obtained. The yields of these particles are compared with the theoretical predictions and the data of other experiments. The measured cross section of the total open charm production ($\sigma _{\mathrm {tot}}(c\overline c)$ = 7.1 $\pm $ 2.3(stat) $\pm $1.4(syst) $\mu $b/nucleon) at the collision c.m. energy $\sqrt {s}$ = 11.8 GeV is well above the QCD model predictions. The contributions of different species of charmed particles to the total cross section of the open charm production in proton-nucleus interactions vary with energy.}

\FullConference{38th International Conference on High Energy Physics\\
                  3-10 August 2016\\
                 Chicago, USA}

\begin{document}

\section{Monte Carlo simulation and selection of events with the charmed particles}
The SERP-E-184 experiment "Investigation of mechanisms of the production of charmed particles in $p$A- interactions at 70 GeV and their decays" at the U-70 accelerator (IHEP, Protvino) was carried out at the SVD-2 (Spectrometer with Vertex Detector) setup \cite{SVD}. This setup was constructed to study the charmed particles production in $pp$- and $p$A-interactions by the SVD collaboration including IHEP (Protvino), JINR (Dubna) and SINP MSU (Moscow). The main elements of the setup are the high-precision micro-strip vertex detector (MSVD) with an active target (AT) and a magnetic spectrometer. The AT contains 5 silicon detectors each 300-$\mu $m thickness and 1-mm pitch strips, a Pb-plate (220 $\mu $m thick) and a C-plate (500 $\mu $m thick), placed as Si-Si-Pb-Si-C-Si-Si. The tracking part of MSVD consists of 10 Si-detectors: four XY pair and one XYUV quadruplet , U and V are the oblique planes. The angular acceptance of MSVD is $\pm $ 250 mrad. The spectrometer features allow one to get the effective mass resolution of $\sigma $ = 4.4 MeV/$c^2$ for $K^0_s$ and 1.6 MeV/$c^2$ for $\Lambda ^0_c$ masses.

Monte Carlo (MC) events were obtained with FRITIOF separately for interactions on C, Si, and Pb with the charm production. Decays of unstable particles happened later within GEANT code. Certain decay modes were imposed for charmed particles ($D^0 \to K^-\pi ^+$, $\overline D^0 \to  K^+\pi ^-$, $D^+ \to K^-\pi ^+\pi ^+$, $D^- \to K^+\pi ^-\pi ^-$, $\Lambda _c^+ \to pK^-\pi ^+$). GEANT3.21 package was used to simulate registration of $p$A-interactions. We analysed the simulated events in order to work out the selection criteria \cite{D0} for $D^0 \to K^-\pi ^+$ and $\overline D^0 \to  K^+\pi ^-$. The effective mass spectra of the $K\pi $ system after applying of all criteria were fitted by the sum of the straight line and the Gaussian function. It gives 1861 $\pm $ 7 MeV/$c^2$ for $D^0$ ($\overline D^0$) mass and the signal-to-noise ratio of (51$\pm $17)/(38$\pm $13). The detection efficiency of ($D^0/\overline D^0$) particles with taking into account of the efficiency of visual inspection is equal to $\epsilon (D^0/\overline D^0$) = 0.036.

For reconstruction of the charged charmed mesons, we analysed the $K\pi \pi $-systems: $D^+ \to K^-\pi ^+\pi ^+$, $D^- \to K^+\pi ^-\pi ^-$. The charged charmed mesons were found by analysing of the events with a three-prong secondary vertexes (the selection procedure is described in \cite{Dplus}). After parametrisation of the spectrum as sum of the Gaussian function and polynomial we were got 15.5 $\pm $ 5.6 (15.0 $\pm $ 4.7) signal events from $D^+$ ($D^-$) meson decay over the background of 16.6 $\pm $ 6.0 (8.7 $\pm $ 2.7) events. Also, the  mass of $D^+$: $M$($D^+$) = 1874 $\pm $ 5 MeV/$c^2$ (PDG $-$ 1869.6), $\epsilon $($D^+$) = 0.014 (efficiency of a signal extraction); the mass of $D^-$: $M$($D^-$)= 1864 $\pm $ 8 MeV/$c^2$,  $\epsilon $($D^-$) = 0.008.

The charmed $\Lambda _c^+ $-baryon was analysed with the three-prong decay $\Lambda _c^+  \to pK^-\pi ^+$. Application of all the selection criteria \cite{Lc} resulted in the effective mass spectrum with the signal-to-noise ratio: (21.6 $\pm $ 6.0)/(16.4 $\pm $ 4) and mass $M$($\Lambda _c^+ $) = 2287 $\pm $ 4 MeV/$c^2$ (PDG $-$ 2286.5), $\epsilon $($\Lambda _c^+$) = 0.011. 
\section{Cross sections for charmed particle production and their A- dependence}
We have calculated inclusive cross sections for charmed particle $i$ 
($i$ = $D^0$, $\overline D^0$, 
$D^\pm $ or $\Lambda _c^+$) using the relation: 

\begin{figure}[]
\begin{minipage}[h]{0.4\linewidth}
\center{\includegraphics[width=1\linewidth]{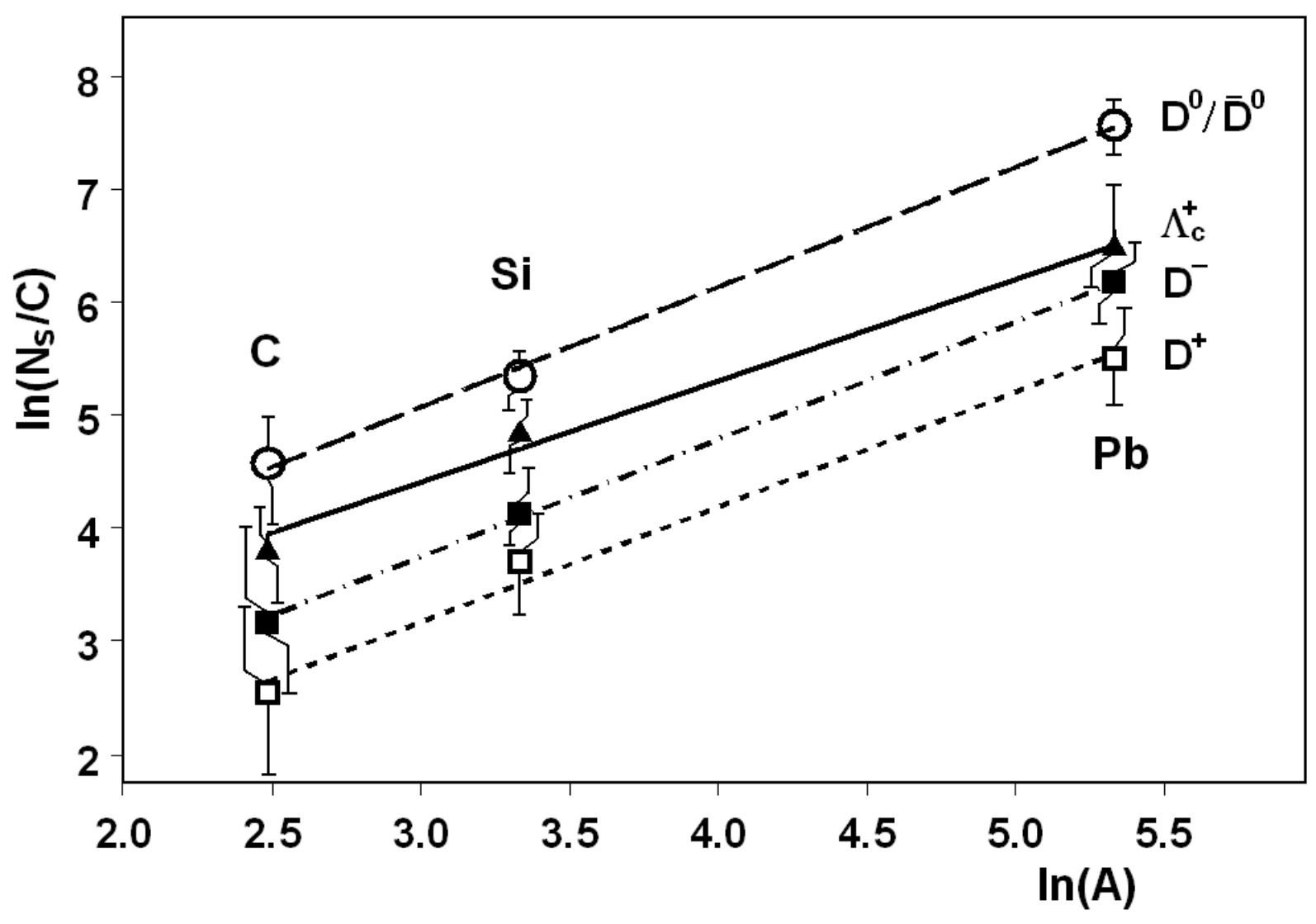}} $a$\\
\end{minipage}
\hfill
\begin{minipage}[h]{0.4\linewidth}
\center{\includegraphics[width=1\linewidth]{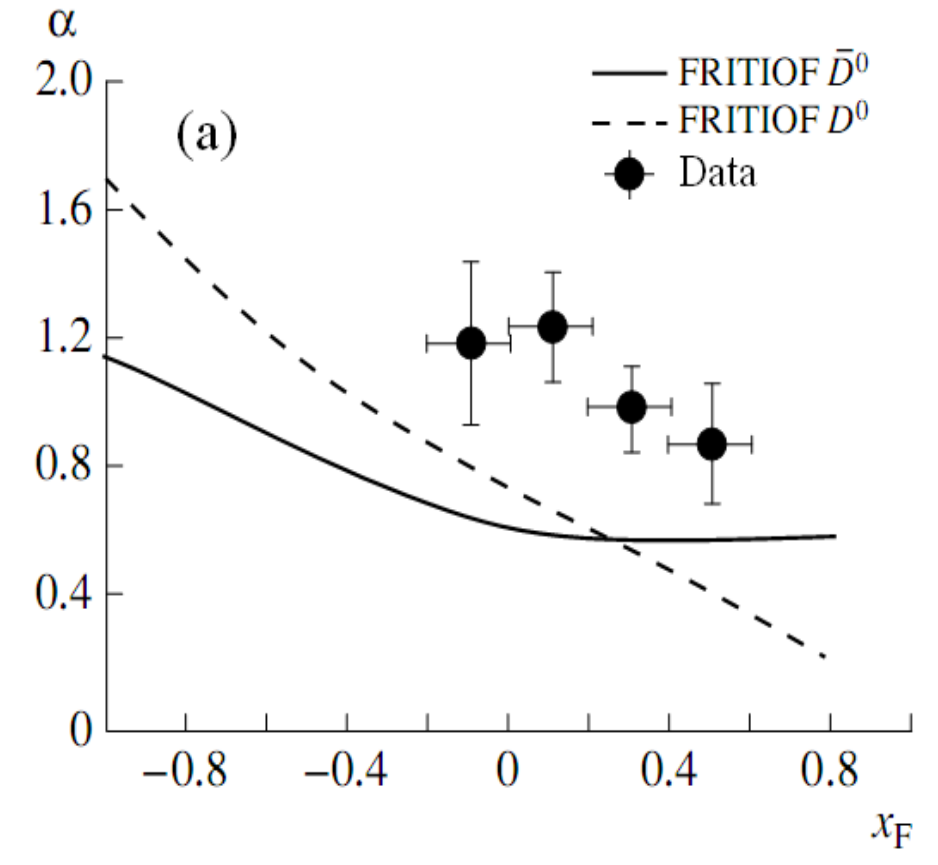}} $b$\\
\end{minipage}
\vfill
\begin{minipage}[h]{0.4\linewidth}
\center{\includegraphics[width=1\linewidth]{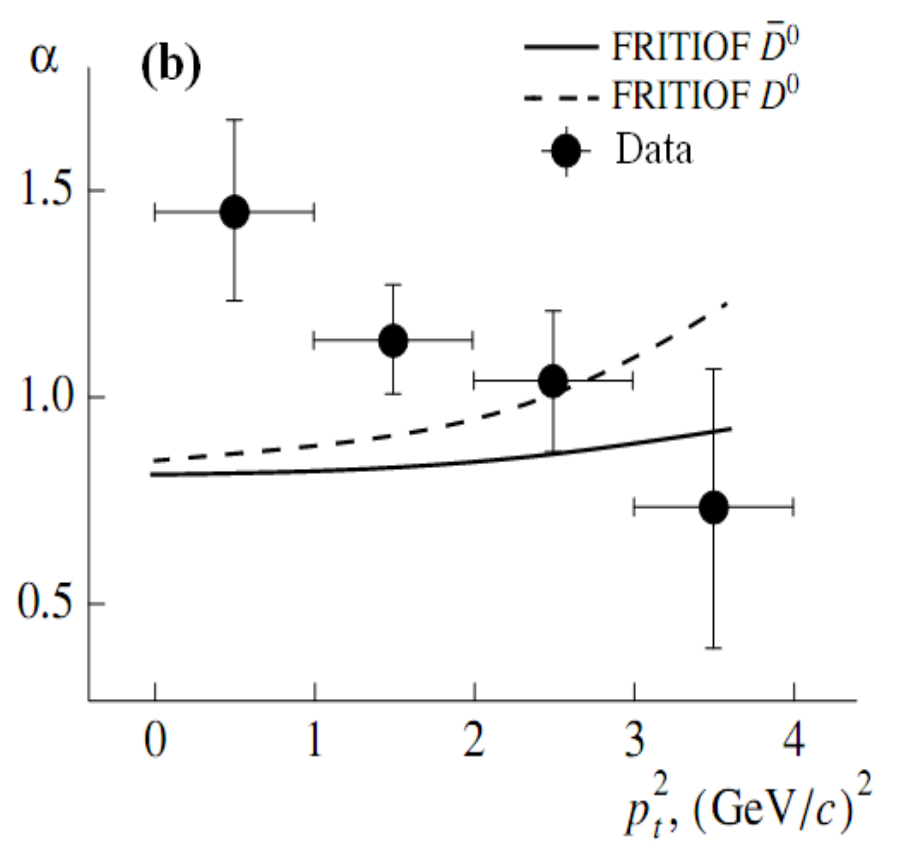}} $c$\\
\end{minipage}
\begin{minipage}[h]{0.6\linewidth}
\center{\includegraphics[width=1\linewidth]{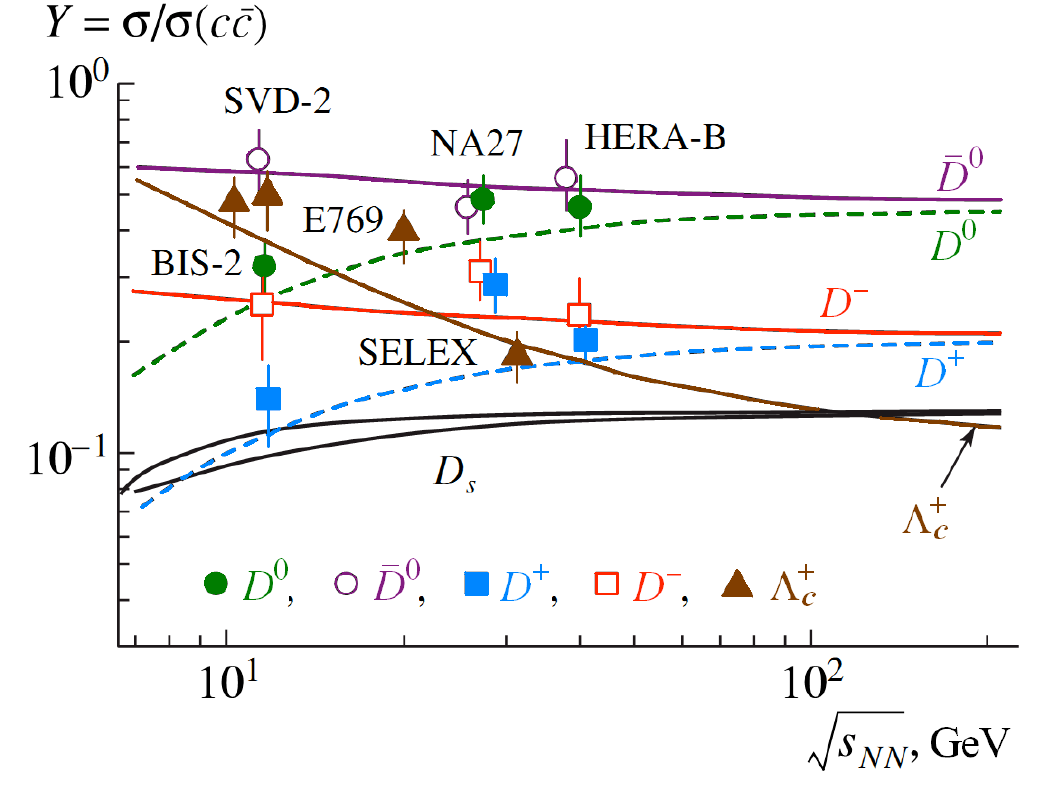}} $d$\\
\end{minipage}
\caption{ ($a$) the A-dependence of cross sections for the charmed particles production in $p$A-interactions; 
the $\alpha $-parameter as a function of ($b$) $x_{\mathrm {F}}$ and ($c$) $p_{\mathrm {T}}^2$ for ($D^0$/$\overline D^0$) particle, the lines describe MC events (FRITIOF); 
($d$) relative yields of charmed particles:  
$\bullet $ $-$ $D^0$, $\circ $ $-$ $\overline D^0$, $\blacksquare$ $-$ 
$D^+$, $\square $ $-$ $D^+$, $\blacktriangle $ $-$ $\Lambda ^+_c$ 
\cite{D0,Dplus,Lc}, the theoretical curves 
(with designation of a particle) are taken from \cite{data}
.}
\label{fig:1}  
\end{figure}

$N_s$($i$) = ($N_0~\sigma $($i$)~A$^\alpha $/($\sigma _{\mathrm {pp}}$ A$^{0.7}$))($B$($i$) 
$\epsilon $($i$))/$K_{\mathrm {tr}}$ or $ln$($N_{\mathrm {s}}$($i$)/$C$($i$)) = 
$\alpha \times ln$(A) + $ln \sigma $($i$), where $C$($i$) = [$N_0/(\sigma _{pp} \times A^{0.7})] \times
[(B(i) \times \epsilon (i))/K_{tr}]$,
$N_s$($i$) determines the number of events in the signal for the $i$-charmed particle produced in the given target, $N_0$ $-$ the number of inelastic interactions in this target, $\sigma $($i$) $-$ the cross section for charmed particle production at a single nucleon of the target.                   
A-dependence of the charmed particle production in $p$A-interactions at the AT (C, Si and Pb) is close to 1 for all charmed particles \cite{Lc} as shown in Fig.~\ref{fig:1}, $a$. For the largest number of the reconstructed mesons ($D^0$/$\overline D$$^0$) the dependences of $\alpha-$parameter on $x_{\mathrm {F}}$ and $p_{\mathrm {T}}^2$ is shown in Fig. \ref{fig:1}, b and c, respectively.  The lines describe MC events (FRITIOF). 
 Relative yields of charmed particles are shown in Fig. \ref{fig:1}, $d$ where  $\bullet $ $-$ $D^0$, $\circ $ $-$ $\overline D^0$, $\blacksquare$ $-$ $D^+$, $\square $ $-$ $D^+$, $\blacktriangle $ $-$ $\Lambda ^+_c$ \cite{D0,Dplus,Lc} are the experimental points, the theoretical curves (with designation of a particle) are taken from \cite{data}.

The total cross section of the charmed particle  production in $pp$ at 70 GeV/c is estimated as
$\sigma _{\mathrm {tot}}$($c\overline c$) = 7.1 $\pm $ 2.3 (stat) $\pm $ 1.4 (syst) $\mu $b/nucleon  \cite{Lc}.

\begin{figure}
\resizebox{\textwidth}{!}{%
  \includegraphics[width=0.34\textwidth]{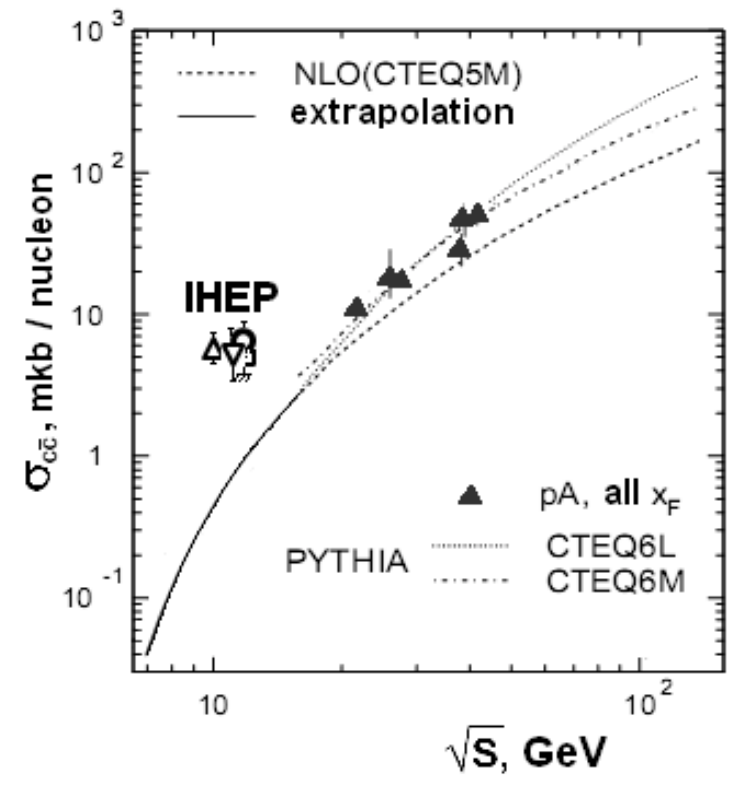}
  \includegraphics[width=0.34\textwidth]{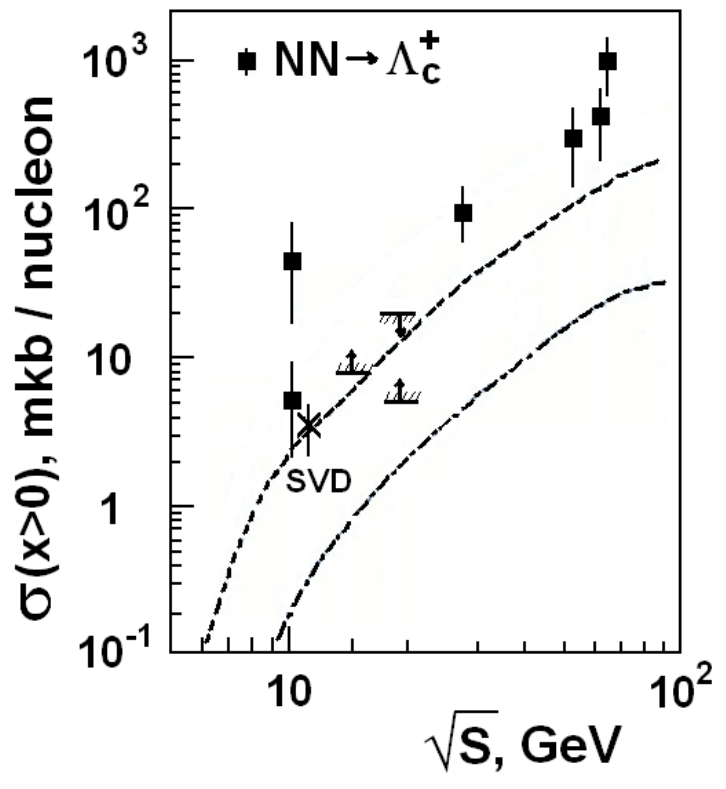}
}
\caption{
Left panel: 
$\sigma _{\mathrm {tot}}$ ($c\overline c$) in $p$A-interactions. Extrapolation  is solid line. 
Experiments: 
$\circ $ $-$ SVD-2, $\square $~$-$ SCAT bubble chamber, $\triangledown $ $-$ beam-dump, 
$\vartriangle $ $-$ BIS-2 spectrometer \cite{data}, other lines are taken from various models (see refs.
in \cite{Lc}.
Right panel:
$\sigma $($\Lambda _c^+$) at $x_\mathrm {F}$  > 0, where $\blacksquare$  $-$ world data, 
X $-$ the result of our experiment, lines -  the model predictions based on QCD \cite{Adep}.  
}
\label{fig:2}       
\end{figure}

\section{Conclusion}
Our basic results of study of the charmed particle production are careful measurements of

$\star $ $\sigma _{\mathrm {tot}}$($c\overline c$) = 7.1 $\pm $ 2.3 (stat) $\pm $ 1.4 (syst) $\mu $b/nucleon 
at c.m. energy $\sqrt s$ = 11.8 GeV that is much above the QCD model predictions (Fig. \ref{fig:2}, the  left panel);

$\star $ the contributions of $\sigma $($i$), where $i$ = $D^0$, $\overline D$$^0$, $D^+$, $D^-$ and 
$\Lambda ^+_c$ into the total cross section $\sigma $($c\overline c$)  
vary at lower collision energies (Fig. \ref{fig:1}, $d$);

$\star $ the cross section for $\Lambda ^+_c$ production at $\sqrt s$ > 30 GeV contradicts $\sigma $($c\overline c$) 
for the open charm production cross section (Fig. \ref{fig:2}, the right panel). $\sigma $($\Lambda ^+_c$) 
are extraordinarily large in this area.

\end{document}